\documentclass[superscriptaddress,nofootinbib,nobibnotes,amsmath,amssymb, twocolumn
, prb
]{revtex4-2}
\usepackage[normalem]{ulem}
\usepackage{siunitx}
\usepackage{graphicx}
\usepackage{dcolumn}
\usepackage{bm}
\usepackage{bbm}
\usepackage{amsmath}
\usepackage{color}
\usepackage[hidelinks]{hyperref}
\usepackage{commath}
\usepackage{tabularx}
\usepackage{booktabs}

\usepackage{bbold}

\definecolor{orange}{rgb}{1,0.5,0}
\begin{document}


\title{ Comment on
``Electron spin resonance and collective excitations in magic-angle twisted bilayer graphene". }
\thanks{mprada@physnet.uni-hamburg.de}
\thanks{\linebreak latieman@physnet.uni-hamburg.de.}

\author{Lars Tiemann}
\thanks{These authors contributed equally.}
\affiliation{%
   Center for Hybrid Nanostructures (CHyN),  Universit\"at Hamburg,   Luruper Chaussee 149, Hamburg 22761 Germany.
}
\author{M. Prada }%
\thanks{These authors contributed equally.}
\affiliation{%
	I. Institute for Theoretical Physics, Universit\"at  Hamburg 
D-22761 Hamburg, Germany
}
\author{Vincent Strenzke}%
\affiliation{%
   Center for Hybrid Nanostructures (CHyN),  Universit\"at Hamburg,   Luruper Chaussee 149, Hamburg 22761 Germany.
}
\author{Robert Blick}
\affiliation{%
   Center for Hybrid Nanostructures (CHyN),  Universit\"at Hamburg,   Luruper Chaussee 149, Hamburg 22761 Germany.
}
\date{\today}
\begin{abstract}
This comment pertains the recent manuscript by Morissette {\it et al.} 
\url{https://arxiv.org/abs/2206.08354v1}. 
The authors 
claim to have found signatures of collective excitations in electron spin resonance experiments that would be 
linked to the correlated structure of magic angle bilayer graphene. 
However, identical resonance features have already been reported in previous works on mono- and few-layer graphene, voiding 
their theoretical framework. 
A straight forward  theoretical picture within the single-particle topologically non-trivial band structure of graphene 
delivers satisfactory explanations for the observation of the resonant features and applies as well to the data presented by  Morissette 
{\it et al.}. %
However, this intuitive picture has been disregarded by the authors. 
\end{abstract}
\maketitle
{\it Introduction: }
The recent work by Morissette {\it et al.} \cite{morissette} reports on resistively detected electron spin resonance (RD-ESR), 
where it is claimed that collective excitations linked to the correlated states in magic angle bilayer graphene (MABG) 
cause particular resonance features. 
However, previous work in graphene mono- and few-layers \cite{jonas,udai,vincent,anlauf,mani} observe similar signatures in RD-ESR.  
The origin of these lines has been reported and is accepted within a single-particle description of spin bands. 
The existing theoretical framework can explain all the experiments consistently, whereas the correlated MABG 
is inconsistent with  the experimental results mentioned above \cite{jonas,udai,vincent,anlauf,mani}. 
The aim of this comment is to present evidence that the experimental data in reference \cite{morissette} simply
confirm and reproduce the existing literature, {which incidentally was not cited in their work. }

\begin{figure*}[!hbt]
\centering
       \includegraphics[width=.99\textwidth]{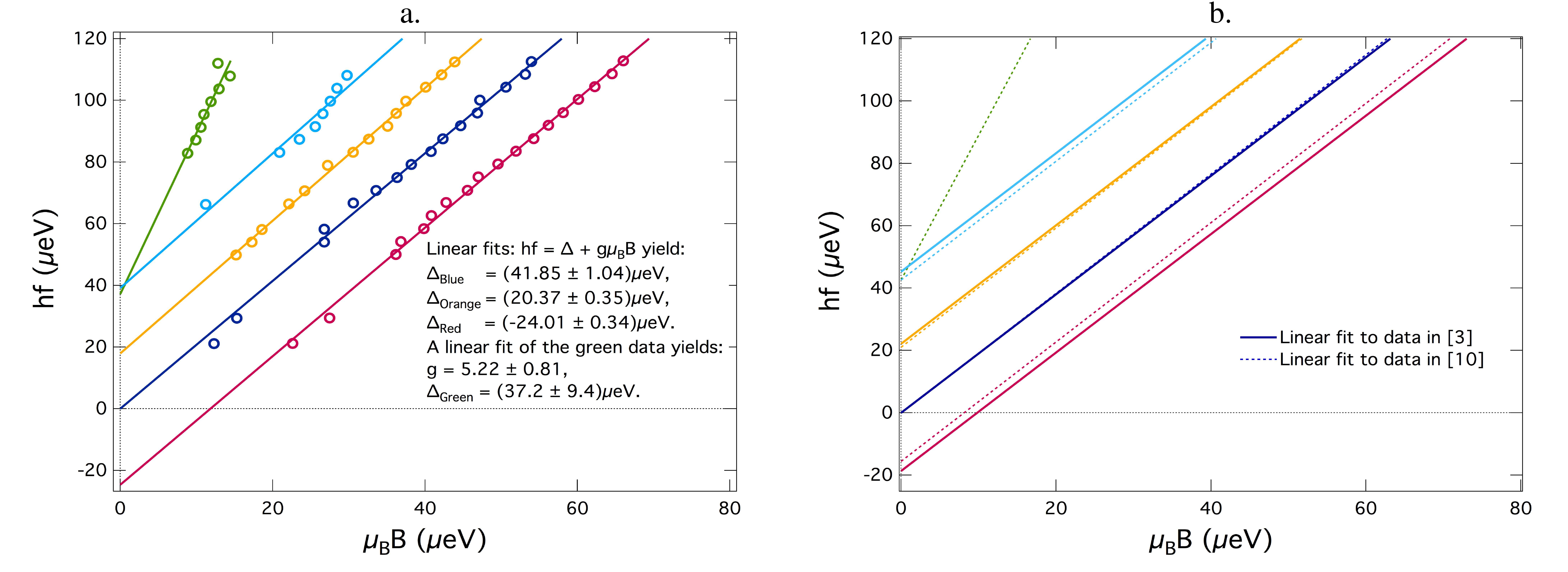}
        \caption[]{ a. Data points  extracted from Fig. 3(b) of Ref. \cite{morissette} with corresponding linear fits (solid lines). 
A zero-field splitting consistent with the 
previously reported value of 42$\mu$eV  can be obtained with the two upper lines (green and light blue), yielding, respectively, 
$\Delta_{\rm Blue}$ and $\Delta_{\rm Orange}$.  
The $g$-factor is extracted from the Zeeman line that crosses the origin (dark blue). 
An additional line with $g$ = 5.2 can be extracted (green line), which compares with the data obtained by Strenzke
\cite{MAVincent} (green dashed lines of b.) with  a $g$-factor of 4.66. 
b. The dashed lines correspond to the 
linear fits to the data previously reported by Strenzke {\it et al.} in $^{13}$C 
graphene \cite{MAVincent,vincent}, whereas the solid lines correspond to 
Singh {\it et al.} \cite{udai}.
        }
\label{fig}
\end{figure*}

{\it Summary of experimental observations: }
RD-ESR is a spin-sensitive probing technique that couples carriers of opposite spin by microwave excitation and detects the response 
resistively \cite{tim,jonas, mani}. In a nutshell, a signal in the longitudinal resistance of a Hall bar is expected whenever 
a resonance condition is met, typically in the frequency-magnetic field plane. 
Four resonance lines are observed by Morissette  {\it et al.} which have already been reported by Singh  {\it et al.} \cite{udai} and by 
Strenzke {\it et al.} \cite{vincent}, where  a common slope can be identified with the Zeeman term $g\mu_BB$. 
The respective intercepts at zero field have been subject of extensive study: 
The most revealing one, first measured by Mani {\it et al.} and subsequently 
reproduced \cite{jonas,udai,vincent} has been identified with intrinsic spin-orbit coupling (SOC)\cite{jonas}. 
The theoretically predicted \cite{konschuh2010tight} zero-field splitting of 42 $\mu$eV 
has been systematically reproduced, not only in RD-ESR experiments but also employing quantum point contacts \cite{new}. 

{\it Discussion of data in Ref. \cite{morissette}: }
The data of Fig. \ref{fig}a. (dots) were extracted using a customary program from Fig. 3(b) of Ref.  \cite{morissette}, where
the solid lines are the corresponding linear fits. For comparison, Fig.  \ref{fig}b. shows the 
linear fits to the data previously reported by Strenzke \cite{MAVincent,vincent} (dashed) and by Singh {\it et al.} \cite{udai}
(solid).
The intercept of the blue line with $h\nu$ axis  $\Delta_{Blue}$ in Fig.  \ref{fig}a. 
has been identified by Morissette and co-workers with the $k_1$ mode of a magnon \cite{morissette},  
$\Delta_{\rm Blue} = (2\pi a/L_x)^2J$, where $J$ would be the spin stiffness, $a$ the Moir\'e lattice constant and $L_x$ 
the sample dimensions. 
In order to proof the validity of their model, it would be necessary to reproduce the data with a sample of different 
dimensions and deduce a scaling law, but only one sample was measured in their work.  
We could extract a  value of $\Delta_{\rm Blue} =$ (41.85 $\pm$ 1.04)$\mu$eV, which coincides with the previously reported value by Mani {\it et al.} \cite{mani} and Sichau {\it et al.} \cite{jonas} of $\Delta_{\rm ISOC} =$ (42.2 $\pm$ 0.8)$\mu$eV,  
also resolved with RD-ESR and identified with intrinsic SOC gap.  
This value was confirmed in bilayer graphene by Banszerus {\it et al.}\cite{new}.
We thus argue here that the blue line simply reproduces previous data on 
{Zeeman} resonance shifted by intrinsic SOC, and thus,  
$\Delta_{\rm Blue} =\Delta_{\rm ISOC}$.

Under special conditions, two additional lines with finite intercepts (or zero-field splittings) 
were observed by Singh {\it et al.}, owing to sublattice splitting \cite{udai}. 
Two lines with intercepts at $\pm \Delta_\gamma$ can be observed as long 
as $\Delta_\gamma$, which accounts for sublattice splitting, 
does not exceed the intrinsic SOC splitting \cite{udai}. 
These findings were later confirmed by  Strenzke {\it et al.} \cite{vincent} on $^{13}C$ graphene. 
They observed an additional shift of these lines owing to the nuclei-induced field,  in perfect agreement with the theory. 
We argue here that these results by Singh {\it et al.} and Strenzke {\it et al.} are simply reproduced in Ref. \cite{morissette}:
Sublattice splitting results on the orange and red lines of Fig. \ref{fig}a., 
as twisted bilayer graphene can give rise to a small asymmetry between the two sublattices, comparable to 
the CVD-grown samples in Singh {\it et al.} and Strenzke {\it et al.}. 
The asymmetry in the splittings with respect to the Zeeman line that cross the origin in the $\nu-B$ 
plane could be related to MABG, but this would require a systematic study with additional data.

There is an additional line (green data points of Fig. \ref{fig}a.) that features a $g$-factor of about 5.2, 
and is similar to the one reported by Strenzke {\it et al.} in \cite{MAVincent}, resulting in  $g\simeq$ 4.6, which 
remains an open question. 

The similarity of the resonance lines measured by Morissette {\it et al.} and the above mentioned publications points towards a common explanation, where  correlations related to MABG are excluded, as it has been measured in CVD graphene. In addition, we stress that
{the lack of all these publications in the references list of the manuscript by Morissette {\it et al.} is surprising as a simple search on \textsl{Google Scholar} using the keywords ``resistively detected electron spin resonance graphene" yields an abundance of relevant content.}


Finally, we enumerate a few deficiencies in the work by Morissette {\it et al.}: 
(i) The validity of their theory should be confirmed with different sample sizes, that would yield accordingly scaling
zero-field splittings, however they only report data on one sample. Existing literature, however, has shown little variation
of the SOC-induced splitting with sample size. 
(ii) The low-magnetic field regime, labelled as regime `1', lacks of experimental evidence. 
Incidentally, previous measurements at low field have reported resonances incompatible with their theory. 
(iii) The above mentioned resonance lines are insensitive to carrier density, and thus, to filling of Moir\'e bands.
(iv) Morissette {\it et al.} claim to find a $g$-factor for the high energy line to be of 4.0, however, our fits indicate a value  
close to 5.2, which cannot be explained within a 2-magnon picture that they invoke in their manuscript. 

{\it Conclusions:} 
In summary, previous RD-ESR studies found identical resonant features of zero-field splitting in monolayer and few-layer graphene. 
However, these studies were completely disregarded by Morissette {\it et al.} in favor of an inconsistent theory that highlights 
and {hyperbolizes} the correlations of MABG. 
The astounding properties of MABG certainly endorse intensive studies, however, 
the results need to be evaluated with greatest scientific care and contrasted with  existing literature.  

\bibliography{bibfile}

\end{document}